\documentstyle[12pt]{article}

\def\be{\begin{equation}}
\def\ee{\end{equation}}
\def\ben{$$}
\def\een{$$}
\def\ba{\begin{array}{c}}
\def\ea{\end{array}}

\begin{document}

\titlepage
\vspace*{1cm}

\begin{center}{\Large \bf
Solvable ${\cal PT}$ symmetric Hamiltonians
 }\end{center}

\vspace{10mm}

\begin{center}
Miloslav Znojil \vspace{3mm}

\'{U}stav jadern\'e fyziky AV \v{C}R, 250 68 \v{R}e\v{z}, Czech
Republic\\

e-mail: znojil@ujf.cas.cz, \today

\end{center}

\vspace{5mm}

\section*{Abstract}

Within the so called ${\cal PT}$ symmetric version of quantum
mechanics a brief review of the exactly solvable models is given.
Distinction is made between the curved and straight coordinate
lines, between their unbounded (aperiodic) and bounded (periodic)
choices, and between the completely and partially solvable cases.

\vspace{5mm}

\section*{[KEYWORDS]}

\noindent [ ${\cal PT}$ symmetry, Schr\"{o}dinger equation in
complex domain, exact solutions]

 \vskip 0.5cm

\section*{[AMS 1991 Mathematics Subject Classification]}

81P10 81Q05 81R40 47B50 34A20 15A57

\newpage

\section{Introduction}

${\cal PT}$ symmetric quantum mechanics \cite{BBMM} has
independently been proposed and used as a methodical laboratory in
quantum physics by several groups of authors: By Caliceti et al
\cite{CGM} in perturbation theory, by Bessis et al \cite{DB} in
field theory and by Andrianov et al \cite{ACDI} in supersymmetric
context.

This short review will pay attention to the (partially as well as
completely) exactly solvable models within this framework, with
emphasis on the results obtained by the present author.

\section{Complete solvability on curved paths}

One of the first exactly and completely solvable examples of a
${\cal PT}$ symmetric system has been found by Cannata et al
\cite{ACDI} and re-discovered by Bender et al \cite{BBJS} more
than one year later. Its modified Schr\"{o}dinger bound state
problem is defined on certain curved, left-right symmetric
``generalized coordinate" lines in the complex plane.
Mathematically, it is defined via exponential potential and proves
exactly solvable in terms of Bessel functions. Due to its
relationship to a power-law forces in the large exponent limit, it
can be most simply interpreted as a certain smooth and
non-Hermitian ${\cal PT}$ symmetric analogue of the current square
well.

Recently, a double-well counterpart of the latter set of models
has been shown exactly solvable, in terms of Laguerre polynomials,
in ref. \cite{8}. In a way similar to the above ``single well"
example its paths of integration are the same, curved complex
lines again. Their spectra exhibit a puzzling and highly
unexpected feature of certain coupling-dependent re-arrangements
mediated by ``unavoided" crossings at critical points. The
phenomenon reflects the non-Hermiticity of the
Hamiltonian~\cite{Herbst}.

Via a suitable Liouvillean change of the variables in the above
double-well-like differential Schr\"{o}dinger equation one can
immediately obtain another Laguerre-related solvable system with
the potential of the Coulombic single-pole form. In this case the
deformation of the integration path plays the beneficial role of a
natural regularization prescription. At the same time it also
leads to the necessity of working with the complex charges in a
way described in ref. \cite{3}. Also the related spectrum of
energies exhibits certain unexpected features: Positivity, a
coexistence of the growth and decrease with the increasing
coupling strength, etc.

A clear mathematical explanation of behaviour as well as a
consistent and/or possible physical interpretation of the
counterintuitive models of this type remain unclarified up to now.
Much better situation emerges in the case of the integration
curves defined as the left-right-symmetric straight lines.

\section{Straight paths}

Using the language of the so called Kustaanheimo - Steifel
transformation (cf. \cite{Pogosyan} for detailed references) both
the above curvilinear examples can be shown equivalent to the
${\cal PT}$ symmetric harmonic oscillator of ref. \cite{10} with a
centrifugal term regularized by the mere downward complex shift of
the (full) real axis,
 \be \left (-\,\frac{d^2}{dx^2} + x^2 -2ic\,x
+ \frac{\alpha^2-1/4}{(x-ic)^2} \right )
 \, \varphi(x) =
(E+c^2)  \, \varphi(x), \ \ \ \ \ \varphi(x) \in
L_2(-\infty,\infty).
 \label{SE}
 \ee
This oscillator with the Laguerre-polynomial normalizable
solutions
 \ben
\varphi_{(\pm n)}(x) = {\cal N} \cdot (x-ic)^{\pm
\alpha+1/2}e^{-(x-ic)^2/2} \ L^{(\pm \alpha)}_n \left [
 (x-ic)^2
 \right ], \ \ \ \ n = 0, 1, \ldots
\label{waves}
 \een
possesses the non-equidistant spectrum of energies $ E=E_{(\pm
n)}=4n+2 \pm 2 \alpha$ and represents  a certain unperturbed limit
of the quartic oscillator models of Buslaev and Grecchi \cite{BG}.
These authors, unfortunately, did not notice the existence of the
``quasi-even", $_{(-n)}-$signed half of the spectrum. This
omission can be easily corrected. One just introduces a
``two-to-one" isospectrality correspondence between the respective
Hermitian and non-Hermitian anharmonic oscillator models of
ref.~\cite{BG}.

Interpretation of the models ``living" on the straight lines
becomes significantly facilitated by the easier identification and
interpretation of their complex components \cite{12}. Immediate
purely analytic constructions recover, e.g., the existence of
models which are in a one-to-one correspondence to the so called
shape-invariant real forces in one dimension (cf. their
presentation in ref. \cite{7}) and on the half line (their ${\cal
PT}$ symmetric counterparts were described and listed in ref.
\cite{4}).

The situation is reviewed in ref. \cite{2a}. A fully general form
of this type of analytic constructions dates back to the
introduction of the so called Natanzon potentials and, in the
present context, is thoroughly analyzed and described in ref.
\cite{1}.

\section{Models with periodic boundary conditions}

A new and promising development of the ${\cal PT}$ symmetric
considerations has been recently inspired by the study of the two-
and three-particle models \cite{Milos}. The ${\cal PT}$
symmetrization of the Hamiltonians has been again conjectured to
be sufficient for keeping their spectrum real. The related
``weakening of the Hermiticity" finds a natural generalization in
the new context.

Particular attention has been paid to the possible non-Hermitian
generalizations of the well known Calogero model \cite{Calogero}.
In this setting the separability of the underlying partial
differential Schr\"{o}dinger equation in the hyperspherical
coordinates helps us to reduce the problem to the ``hyperangular"
ordinary differential equation defined on a finite interval. In
this way, in the simplest cases one has to solve the complexified
ordinary differential equations of the generalized
P\"{o}schl-Teller type,
 \be
\left ( -\frac{d^2}{d \phi^2} + \frac{\ell(\ell+1)}{\sin^2\phi}
 + \frac{\lambda(\lambda+1)}{\cos^2\phi} \right
) \chi(\phi) = E\,\chi(\phi)
 \label{angular}
 \ee
on an interval $\phi \in (-M\,\pi/2, M\,\pi/2)$ with a suitable
integer $M$. These equations can be solved exactly in terms of the
hypergeometric functions~\cite{Khare}.

The strongly repulsive singularities at $\phi_j=j\,\pi/2$ are
currently not penetrable \cite{Fluegge}. Here, they become
regularized in a ${\cal PT}$ symmetric manner which parallels a
few older constructions on the unbounded intervals \cite{4}. The
quasi-symmetric and quasi-antisymmetric solutions arise from
certain {\em ad hoc} boundary conditions \cite{Milos,comment}.

For the most elementary illustration let us now choose $\lambda=0$
and $M=2$. Then, the differential equation (\ref{angular})
possesses the two independent hypergeometric solutions,
 \ben
\chi^{(\pm)}(\phi)=(\sin\,\phi)^{1/2\pm \alpha} \
_2F_1(u^{(\pm)},v^{(\pm)};1\pm \alpha;\sin^2\phi), \ \ \
\alpha=\ell+1/2>0
 \een
where $2u^{(\pm)}=1/2- \beta \pm \alpha$ and $2v^{(\pm)}=1/2+
\beta \pm \alpha$. On the boundary of convergence $\sin^2 \phi =
1$ the matching of the logarithmic derivatives is equivalent to
the termination of this series to the Gegenbauer polynomials,
 \ben
\chi(\phi)=\chi_{(\pm k)}(\phi)=(\sin\,\phi)^{1/2\pm \alpha} \
C_k^{1/2\pm \alpha}(\cos \,\phi),\ \ \ \ \ \ k = 0, 1, \ldots\ .
 \een
The construction also quantizes the energies and gives them in the
closed form,
 \ben
 E=E_{(\pm k)}=
 (k \pm \alpha + 1/2)^2,\ \ \ \ \ \ k = 0, 1, \ldots\ .
  \een
This set of eigenvalues is composed of the two subsets in a way
which resembles the above-mentioned non-equidistant spectrum of
the  ${\cal PT}$ symmetrized singular harmonic
oscillator~(\ref{SE})~\cite{10}.

\section{Conclusions and outlook}

Historically, one of the first persuasive manifestations of the
merits and power of the ${\cal PT}$ symmetry has been offered by
Bender and Boettcher \cite{BB} who discovered the quasi-exact
(i.e., incomplete) solvability of the most common and popular
quartic polynomial oscillators.

The plausible reasons of the unexpected delay of such an
``obvious" observation are closely related to the above-mentioned
``forgotten" energies. One has to keep in mind the ``spontaneous"
regularity of the singular potentials within the new formalism.
This observation was made explicit in our paper \cite{6} where the
presence of the two additional singular terms has still been shown
compatible with the quasi-exact solvability of quartic potentials.

In the quasi-exact context the changes of variables can play the
same role as in the completely solvable models. This has been
illustrated by the particular constructions of the decadic model
\cite{2} and of the harmonic + Coulomb superposition \cite{11}.
Further work in this direction is in progress \cite{sextic}.


\section*{Acknowledgements}

Research assisted by the Grant Agency of the Academy of Sciences
of the Czech Republic, contract Nr. A 1048004.


\end{document}